\documentclass[twocolumn]{article} % default is 10 pt
\usepackage{graphicx} % needed for including graphics e.g. EPS, PS
\usepackage{amssymb}

\long\def\comment#1{}

\makeatletter
\def\@normalsize{\@setsize\normalsize{10pt}\xpt\@xpt
\abovedisplayskip 10pt plus2pt minus5pt\belowdisplayskip
\abovedisplayskip \abovedisplayshortskip \z@
plus3pt\belowdisplayshortskip 6pt plus3pt
minus3pt\let\@listi\@listI}

\def\subsize{\@setsize\subsize{12pt}\xipt\@xipt}
\def\section{\@startsection {section}{1}{\z@}{1.0ex plus
1ex minus .2ex}{.2ex plus .2ex}{\large\bf}}
\def\subsection{\@startsection
   {subsection}{2}{\z@}{.2ex plus 1ex} {.2ex plus .2ex}{\subsize\bf}}
\makeatother

\begin{document}

% don't want date printed
\date{}

% >>>>>>>>>>>>>>>>>>>>>>>  Put your title here <<<<<<<<<<<<<<<<<<<<<<<<
% make title bold and 14 pt font (Latex default is non-bold, 16pt)
\title{\huge \bf {Stochastic Recovery Of Sparse Signals From Random Measurements}}

% >>>>>>>>>>>>>>>>>>>>>>> Author's Name, Thanks or Affliation <<<<<<<<
\author{M. Andrecut 
 \thanks{Manuscript submitted July 20, 2010. Institute for Space Imaging Science, University of Calgary, 
 2500 University Drive NW, Calgary, Alberta, T2N 1N4, Canada, Email: mandrecu@ucalgary.ca.
 }
}

\maketitle
\thispagestyle{empty}

%\subsection*{}

% >>>>>>>>>>>>>>>>>>>>>>>>> Keywords and Abstract <<<<<<<<<<<<<<<<<<<<<
% Replace with your own keywords and abstract.  Text will be in italics
{\hspace{1pc} {\it{\small Abstract}}{\bf{\small---Sparse signal
recovery from a small number of random measurements is a well known
NP-hard to solve combinatorial optimization problem, with important
applications in signal and image processing. The standard approach
to the sparse signal recovery problem is based on the basis pursuit
method. This approach requires the solution of a large convex optimization
problem, and therefore suffers from high computational complexity.
Here, we discuss a stochastic optimization method, as a low-complexity
alternative to the basis pursuit approach.

}}
 }

% >>>>>>>>>>>>>>>>>>>>>> START OF YOUR PAPER <<<<<<<<<<<<<<<<<<<<<<<<<<<<<<
% Typically paper starts of with an Introduction header.  Replace text in
% the french braces if you see fit.  I also typically name the label the
% same as the section

\section{Introduction}
\label{Introduction}

There has been an increasing interest in the problem of sparse signal
recovery from random measurements, which is strongly motivated by
the recently introduced framework of Compressive Sensing (CS) (see
\cite{key-1}-\cite{key-7} and the references within). The main idea
of CS is that if the signal is compressible, then a small number of
random measurements contain sufficient information for its approximate
or exact recovery. CS has promising applications in signal and image
processing, and it could potentially lead to interesting models for
various interactions performed at the biological level \cite{key-8}.

The problem of sparse
signal recovery from random projections requires the minimization
of the $\ell_{0}$ norm of the candidate solution. This is generally
impossible to solve, since it requires an intractable combinatorial
search. A common alternative is to consider the convex problem, known
as Basis Pursuit (BP), which requires the minimization of the $\ell_{1}$
norm, as a sparsity-promoting functional. Here, we discuss a Stochastic
Optimization (SO) method, which provides a low-complexity alternative
to the standard Basis Pursuit (BP) approach used in CS \cite{key-1}-\cite{key-7}.
The considered SO method has the advantage of a very easy implementation,
comparing to the highly complex BP approach. The objective function
of the SO method is also different, and it is defined as the product
between the $\ell_{1}$ norm and the \textit{spectral entropy} of
the candidate solution. This definition of the objective function
improves the signal reconstruction, since it is equivalent with a
weighted $\ell_{1}$ norm functional, where the weights correspond
to the \textit{self-information} of each component of the signal.
Thus, by using such an objective function, the SO minimization method
focuses on entries where the weights are small, and which by definition
correspond to the non-zero components of the sparse signal.

\section{Compressive Sensing Framework}
\label{Compressive Sensing Framework}

Let us give a short description of the CS framework \cite{key-1}-\cite{key-7}.
Assume that we acquire a discrete signal:
\begin{equation}
z=[z_{1},...,z_{M}]^{T}\in\mathbb{R}^{M},
\end{equation}
and 
\begin{equation}
\Psi=\left\{ \psi^{(m)}|\psi^{(m)}\in\mathbb{R}^{M},m=1,...,M\right\},
\end{equation}
is an orthonormal basis of vectors spanning $\mathbb{R}^{M}$ (here
$T$ stands for transposition of vectors and matrices). We denote by:
\begin{equation}
\hat{\Psi}=[\psi^{(1)}|...|\psi^{(M)}],
\end{equation}
the matrix with the columns given by the basis vectors. Obviously,
the matrix $\hat{\Psi}$ corresponds to a unitary transformation,
i.e. it satisfies the orthogonality condition: 
\begin{equation}
\hat{\Psi}^{T}\hat{\Psi}=\hat{\Psi}\hat{\Psi}^{T}=\hat{I}_{M},
\end{equation}
 where $\hat{I}_{M}$ is the $M\times M$ identity matrix. We say
that $\Psi$ provides a sparse representation of $z$, if $z$ is
well approximated by a linear combination of a small set of vectors
from $\Psi$, i.e. there exists a set of indices $\{m_{1},...,m_{K}\}\subset\{1,...,M\}$,
for small $K\ll M$, such that:\begin{equation}
z=\sum_{k=1}^{K}x_{m_{k}}\psi^{(m_{k})},\; x_{m_{k}}\neq0.\end{equation}
Therefore:
\begin{equation}
\hat{\Psi}^{T}z=x=[x_{1},...,x_{M}]^{T}\in\mathbb{R}^{M},
\end{equation}
is a sparse signal, representing the acquired signal $z\in\mathbb{R}^{M}$
in the basis $\Psi$. Reciprocally, by knowing $x$ one can easily
synthesize $z$ as following: 
\begin{equation}
z=\hat{\Psi}x.
\end{equation}
Now, let us consider a set of vectors: 
\begin{equation}
\Phi=\left\{ \varphi^{(n)}|\varphi^{(n)}\in\mathbb{R}^{M},n=1,...,N\right\},
\end{equation}
and the corresponding matrix: 
\begin{equation}
\hat{\Phi}=[\varphi^{(1)}|...|\varphi^{(N)}],
\end{equation}
such that $N\leq M$. We use these vectors to collect $N$ measurements
of the sparse signal $x$:
\begin{equation}
\hat{\Phi}^{T}x=y=[y_{1},...,y_{N}]^{T}\in\mathbb{R}^{N},
\end{equation}
such that: 
\begin{equation}
y_{n}=\left\langle \varphi^{(n)},x\right\rangle =\sum_{m=1}^{M}\varphi_{m}^{(n)}x_{m},\; n=1,...,N.
\end{equation}
The CS theory shows that it is possible to construct a set of vectors
$\Phi$, such that the measurements $y$ preserve the essential information
about the sparse signal $x$, and therefore the sparse signal $x$
can be recovered from the measurements $y$. More specifically, the
CS theory shows that the matrices $\hat{\Psi}$ and $\hat{\Phi}$
must satisfy the \textit{restricted isometry condition} \cite{key-1}-\cite{key-7}:
\begin{equation}
(1-\delta_{z})\left\Vert z\right\Vert _{2}\leq\left\Vert \hat{\Phi}^{T}\hat{\Psi}^{T}z\right\Vert _{2}\leq(1+\delta_{z})\left\Vert z\right\Vert _{2},
\end{equation}
where $\left\Vert \cdot\right\Vert _{2}$ denotes the $\ell_{2}$
(Euclidean) norm. The restricted isometry constant $\delta_{z}$ is
defined as the smallest constant for which this property holds for
all $K$-sparse vectors $z\in\mathbb{R}^{M}$ in the basis $\Psi$.
The matrices $\hat{\Phi}$ and $\hat{\Psi}$ must also satisfy the
\textit{incoherence condition}, which requires that the rows of $\hat{\Phi}$
cannot sparsely represent the columns of $\hat{\Psi}$ (and vice versa)
\cite{key-1}-\cite{key-7}. It has been shown that one can generate
such a measurement matrix $\hat{\Phi}$ with high probability, if
the elements of the matrix are drawn independently from certain probability
distributions, such as the Gaussian distribution or the Bernoulli
distribution \cite{key-1}-\cite{key-7}. This is a consequence of
the fact that in high dimensions the probability mass of certain random
variables concentrates strongly around their expectation. Also, recent
theoretic considerations have shown that in order to achieve the restricted
isometry condition, any $M\times N$ matrix $\hat{\Phi}$ must have
at least $N\simeq cK\leq M$ columns, $c=c(M,K)\sim\log(M/K)>1$,
in order for the observation $y\in\mathbb{R}^{N}$ to allow an accurate
reconstruction of $x$ \cite{key-1}-\cite{key-7}.

Searching for the sparsest $x$ that matches $y$, subject to the
measurements $\Phi$, leads to the $\ell_{0}$ optimization problem:
\begin{equation}
x=\arg\min_{x\in\mathbb{R}^{M}}\left\Vert x\right\Vert _{0},\; s.t.\;\hat{\Phi}^{T}x=y.
\end{equation}
Here, 
\begin{equation}
\left\Vert x\right\Vert _{0}=\sum_{m=1}^{M}\left[1-\delta(x_{m},0)\right],
\end{equation}
is the $\ell_{0}$ norm, measuring the number of nonzero coefficients
in the vector $x$, and 
\begin{equation}
\delta(a,b)=\left\{ \begin{array}{ccc}
1 & if & a=b\\
0 & if & a\neq b\end{array}\right.,
\end{equation}
is Dirac's function. Unfortunately, this problem is known to be an
NP-hard combinatorial optimization problem, requiring the enumeration
of all possible collections of columns in the matrix $\hat{\Phi}^{T}$
and searching for the smallest collection which best approximates
the signal $y$ \cite{key-1}-\cite{key-7}. The standard approach
to the sparse recovery problem is based on the convexification of
the objective function, obtained by replacing the $\ell_{0}$ norm
with the $\ell_{1}$ norm:
\begin{equation}
\left\Vert x\right\Vert _{1}=\sum_{m=1}^{M}\left|x_{m}\right|.
\end{equation}
The resulting optimization problem:
\begin{equation}
x=\arg\min_{x\in\mathbb{R}^{M}}\left\Vert x\right\Vert _{1},\; s.t.\;\hat{\Phi}^{T}x=y.
\end{equation}
is known as BP, and it can be solved using linear programming techniques
whose computational complexities are polynomial \cite{key-1}-\cite{key-7}.
However, in most real applications the BP approach requires the solution
of a very large convex, non-quadratic optimization problem, and therefore
suffers from high computational complexity \cite{key-1}-\cite{key-7}.

A summarizing diagram of the CS framework is given in Figure 1. During
the encoding process the signal $z\in\mathbb{R}^{M}$ is first transformed
into a sparse signal $x\in\mathbb{R}^{M}$, using the $M\times M$
orthogonal transform $\hat{\Psi}^{T}$ (a typical situation corresponds
to Fourier and wavelet representations of natural images). In the
next step the sparse signal $x\in\mathbb{R}^{M}$ is compressed into
$y\in\mathbb{R}^{N}$, using the $N\times M$ random projection matrix
$\hat{\Phi}^{T}$, with $N<M$. The decoding process requires only
the compressed vector $y\in\mathbb{R}^{N}$ and the matrices $\hat{\Phi}$
and $\hat{\Psi}$, and consists also in two steps. In the first step,
the sparse signal $x\in\mathbb{R}^{M}$ is recovered by solving the
$\ell_{1}$ minimization problem. In the second step the original
signal $z\in\mathbb{R}^{M}$ is synthesized from $x$, using the orthogonal
transform $\hat{\Psi}$.

\begin{figure*}
\centering
\includegraphics[scale=0.7]{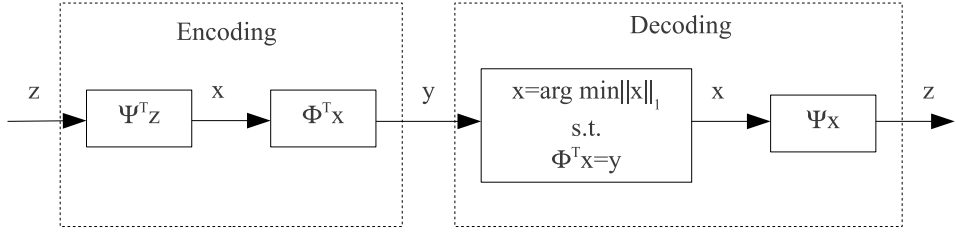}
\caption{Compressive sensing framework.}
\label{Fig1}
\end{figure*}

\section{Stochastic Optimization Approach}
\label{Stochastic Optimization Approach}

Our SO approach is based on the observation formulated in the following
theorem:

\textbf{\textit{Theorem:}} The solution $x$ of the BP problem has
the following form:\begin{equation}
x=x'+\xi,\end{equation}
where $x'$ is a solution of the underdetermined linear system:
\begin{equation}
\hat{\Phi}^{T}x'=y,
\end{equation}
and $\xi$ is an unknown vector from the null subspace of the measurement
matrix $\hat{\Phi}^{T}$.

\textbf{\textit{Proof:}}
Let us assume that $\xi$ is a vector from the null subspace of the
measurement matrix $\hat{\Phi}^{T}$, i.e. it can be written as a
linear combination:
\begin{equation}
\xi=\sum_{m=1}^{M}a_{m}q^{(m)},
\end{equation}
where $a_{m}\in\mathbb{R}$, and $q^{(m)}$ are the columns of the
null subspace projection operator:
\begin{equation}
\hat{Q}=[q^{(1)}|...|q^{(m)}]=\hat{I}_{M}-\hat{\Phi}\left(\hat{\Phi}^{T}\hat{\Phi}\right)^{-1}\hat{\Phi}^{T}.
\end{equation}
Obviously we have: 
\begin{equation}
\hat{\Phi}^{T}\hat{Q}=0\Leftrightarrow\hat{\Phi}^{T}q^{(m)}=0,\; m=1,..,M,
\end{equation}
and consequently:
\begin{equation}
\hat{\Phi}^{T}x=\hat{\Phi}^{T}\left(x'+\xi\right)=\hat{\Phi}^{T}x'=y.
\end{equation}
Thus, $x'$ must be a solution of the underdetermined linear system:
$\hat{\Phi}^{T}x'=y$.

Reciprocally, let us assume that $x'$ is a solution of the underdetermined
linear system: $\hat{\Phi}^{T}x'=y$. Thus, since $x=x'+\xi$ is the
solution of the BP problem, then it also satisfies the constraint
$\hat{\Phi}^{T}x=y$, which implies $\hat{\Phi}^{T}\xi=0$, and therefore
$\xi$ must be a vector from the null subspace of $\hat{\Phi}^{T}$
$\blacksquare$.

A good initial solution (with minimum $\ell_{2}$ norm) of the underdetermined
linear system $\hat{\Phi}^{T}x'=y$, is given by:
\begin{equation}
x'=\Phi^{\dagger}y,\end{equation}
where\begin{equation}
\Phi^{\dagger}=\Phi\left(\hat{\Phi}^{T}\hat{\Phi}\right)^{-1},
\end{equation}
is the Moore-Penrose pseudo-inverse \cite{key-9}, since obviously
we have:
\begin{equation}
\hat{\Phi}^{T}x'=\hat{\Phi}^{T}\Phi^{\dagger}y=\hat{\Phi}^{T}\Phi\left(\hat{\Phi}^{T}\hat{\Phi}\right)^{-1}y=\hat{I}_{N}y=y.
\end{equation}
Unfortunately, it is hard to find the unique vector
$\xi=x-x'$ from the null subspace, which corresponds to the unique
BP solution $x$. This is where we use the stochastic search approach,
in order to avoid the difficult to implement deterministic optimization
methods. 

Our approach is inspired by the Threshold Accepting (TA)
method \cite{key-10}. TA method is a deterministic analog to the well
known Simulated Annealing (SA) method \cite{key-11}. It is a refined
search procedure which escapes local minima by accepting solutions
which are not worse by more than a given threshold. The algorithm
is deterministic in the sense that we fix a number of iterations and
explore the neighborhood with a fixed number of random steps during
each iteration. Analogously to the SA method, the threshold is decreased
successively and approaches zero in the last iteration.
The main difference between SA and TA is that for SA the threshold
is modeled as a random variable, while for TA the threshold is deterministic.
The TA algorithm has the advantage of an easy parametrization, it
is robust to changes in problem characteristics and works well for
many hard problems.

In our case, the initial solution $x\leftarrow x'$ is randomly perturbed in the
null subspace of the matrix $\hat{\Phi}^{T}$, by adding a small random
contribution of the column $q^{(m)}$, $m=1,...,M$, of the matrix
$\hat{Q}$: 
\begin{equation}
\widetilde{x}=x'+\sigma\alpha q^{(m)}.
\end{equation}
Here $\alpha>0$ gives the magnitude of the perturbation, and $\sigma$
is a uniform binary random variable, which provides the sign of the
perturbation, $\sigma\in\{\pm1\}$. Obviously, at every random transition
step, the new candidate solution $\widetilde{x}$ still satisfies
the linear constraint:
\begin{equation}
\hat{\Phi}^{T}\widetilde{x}=\hat{\Phi}^{T}\left(x'+\alpha\sigma q^{(m)}\right)=\hat{\Phi}^{T}x'=y,
\end{equation}
since $\hat{\Phi}^{T}q^{(m)}=0$. Thus, in the framework of $\ell_{1}$
norm minimization, the new random candidate solution $\widetilde{x}$
is accepted ($x\leftarrow\widetilde{x}$) if:
\begin{equation}
\left\Vert \widetilde{x}\right\Vert _{1}-\left\Vert x\right\Vert _{1}\leq\theta,
\end{equation}
and the process continues with a new perturbation. 

We assume that the threshold parameter $\theta\geq0$ and the magnitude
of the perturbations $\alpha\geq0$, follow an exponential
schedule, decreasing by a fixed factor $\lambda\in(0,1)$: 
\begin{equation}
\theta\leftarrow\lambda\theta,\quad \alpha\leftarrow\lambda\alpha, 
\end{equation}
at each iteration. If $\theta_{i}$
and $\theta_{f}$ are the initial and, respectively the final values
of the threshold, then we set $\lambda=(\theta_{f}/\theta_{i})^{1/T}$,
where $T$ is the total number of iterations. 

While this approach
provides a relatively simple SO method for $\ell_{1}$ norm minimization,
it still can be improved by accelerating its convergence. This can
be done by replacing the standard $\ell_{1}$ norm with a weighted
$\ell_{1}$ norm, which penalizes the small components of the candidate
solution. The weighted $\ell_{1}$ norm can be derived using the concept
of \textit{spectral entropy}, as shown below.

We assume that $x=[x_{1},...,x_{M}]^{T}$ is the \textit{spectral
decomposition} of some discrete signal $z\in\mathbb{R}^{M}$ in the
basis $\Psi$: 
\begin{equation}
z=\hat{\Psi}x=\sum_{m=1}^{M}x_{m}\psi^{(m)}\in\mathbb{R}^{M},
\end{equation}
Thus, $\left|x_{m}\right|$ represents the \textit{spectral amplitude}
of the component $\psi^{(m)}$, in the spectral decomposition of $z\in\mathbb{R}^{M}$.

We define the \textit{spectral signature} of $z\in\mathbb{R}^{M}$
in the basis $\Psi$ as following:
\begin{equation}
p(x,\Psi)=[p_{1},...,p_{M}]^{T},
\end{equation}
where
\begin{equation}
p_{m}=\frac{\left|x_{m}\right|}{\sum_{i=1}^{M}\left|x_{i}\right|}=\frac{\left|x_{m}\right|}{\left\Vert x\right\Vert _{1}},\; m=1,...,M.
\end{equation}

We also define a probability measure $P$ for $x$ by:
\begin{equation}
P(x_{m},\psi^{(m)})=p_{m},
\end{equation}
which means that $p_{m}$ is the probability that the spectral decomposition
of $z\in\mathbb{R}^{M}$ in the basis $\Psi$, has the component $\psi^{(m)}$
with an amplitude $\left|x_{m}\right|$. Obviously, we have:
\begin{equation}
\sum_{m=1}^{M}p_{m}=1,
\end{equation}
and $p$ is a probability distribution, associated with $x$. 

Therefore, $x$ can be modeled as a random variable in the probability space
$(\mathbb{R}^{M},\Psi,P)$. Also, $x$ can be viewed as an \textit{information
source}, with its statistics defined by $p$.

Since $x\in(\mathbb{R}^{M},\Psi,P)$ may be viewed as an information
source, we can further define its \textit{self-information} provided
by the component $\psi^{(m)}$ as following \cite{key-12,key-13}:
\begin{equation}
h(x_{m},\psi^{(m)})=-\log_{M}P(x_{m},\psi^{(m)})>0.
\end{equation}

The average self-information, over all components of the basis $\Psi$,
defines the \textit{spectral entropy} of the source $x\in(\mathbb{R}^{M},\Psi,P)$
\cite{key-12,key-13}:
\begin{equation}
H(x,\Psi)=\sum_{m=1}^{M}P(x_{m},\psi^{(m)})h(x_{m},\psi^{(m)}).
\end{equation}
In the case of $p_{m}=0$, we consider $p_{m}\log_{M}p_{m}\equiv0$,
which is consistent with the limit: $\lim_{p\rightarrow0^{+}}p\log_{M}p=0$.

Obviously, we have $H(x,\Psi)\in[0,1]$. A high value $H(x,\Psi)\simeq1$
means that the source $x\in(\mathbb{R}^{M},\Psi,P)$ is just noise,
i.e. all the components $\psi^{(m)}$ have equiprobable amplitudes,
while a smaller value $0<H(x,\Psi)\ll1$, means that the source has
some intrinsic structure (or order), i.e. some components have a stronger
amplitude than others. Equivalently, if $x\in\mathbb{R}^{M}$ is a
sparse signal, then its entropy will be low, while if $x$ is not
sparse then its entropy will be high. 

Now, let us consider the following functional, defined as the product
between the spectral entropy and the $\ell_{1}$ norm:
\begin{equation}
F(x,\Psi)=\left\Vert x\right\Vert _{1}H(x,\Psi).
\end{equation}
This functional corresponds to the weighted $\ell_{1}$ norm of the vector $x\in\mathbb{R}^{M}$:
\begin{equation}
\left\Vert x\right\Vert _{w1}=F(x,\Psi)=\sum_{m=1}^{M}w_{m}\left|x_{m}\right|,\end{equation}
where the weights are given by the self-information of each component:
\begin{equation}
w_{m}=h(x_{m},\psi^{(m)}).
\end{equation}

By replacing the standard $\ell_{1}$ norm, $\left\Vert x\right\Vert _{1}$,
with its weighted version, $\left\Vert x\right\Vert _{w1}$, one penalizes
the small components, $x_{m}$, of the candidate solution, since:
\begin{equation}
w_{m}=\lim_{\left|x_{m}\right|\rightarrow0}\log_{M}\frac{\left\Vert x\right\Vert _{1}}{\left|x_{m}\right|}=\infty.
\end{equation}
Thus, in the stochastic iterative process, the solution will concentrate
around the small weights, which correspond to the non-zero components
of the sparse signal. 

In order to avoid the singularity in the weight
estimation, we introduce a small parameter $0<\varepsilon\ll1/M$,
such that:
\begin{equation}
w_{m}=\log_{M}\frac{\left\Vert x\right\Vert _{1}+M\varepsilon}{\left|x_{m}\right|+\varepsilon}>0.
\end{equation}

Therefore, this analysis shows that from the point of view of solving
the sparse signal recovery problem we need to find the source $x\in(\mathbb{R}^{M},\Psi,P)$
which minimizes the weighted $\ell_{1}$ norm, subject to a linear
constraint system, imposed by the random measurements: 
\begin{equation}
x=\arg\min_{x\in\mathbb{R}^{M}}\left\Vert x\right\Vert _{w1},\; s.t.\;\hat{\Phi}^{T}x=y.
\end{equation}
Using the same threshold accepting approach as before, a new candidate
solution $\widetilde{x}$ is accepted ($x\leftarrow\widetilde{x}$)
if:
\begin{equation}
\left\Vert \widetilde{x}\right\Vert _{w1}-\left\Vert x\right\Vert _{w1}\leq\theta.
\end{equation}
Thus, the pseudo-code of the proposed SO algorithm can formulated as following:\bigskip{}

\# SO sparse signal recovery algorithm: \\*
\# Initialize the parameters \\*
$\theta_{i}$, $\theta_{f}$ , $\alpha_{i}$, $T$; \\*
\# Compute $\lambda$ \\*
$\lambda=(\theta_{f}/\theta_{i})^{1/T}$; \\*
\# Initialize the solution \\*
$x\leftarrow\hat{\Phi}^{\dagger}y$; \\*
$F\leftarrow\left\Vert x\right\Vert _{w1}$; \\*
\# Compute the null subspace projection operator \\*
$\hat{Q}_{\bot}\leftarrow\hat{I}_{M}-\hat{\Phi}\left(\hat{\Phi}^{T}\hat{\Phi}\right)^{-1}\hat{\Phi}^{T}$; \\*
\# Set the initial parameters \\*
$\theta\leftarrow\theta_{i}$; $\alpha\leftarrow\alpha_{i}$; \\*
FOR($t=1,...,T$)\{ \\*
FOR($m=1,...,M$)\{ \\*
\#Compute a candidate solution \\*
$\sigma\leftarrow sign(rnd(1)-0.5)$; \\*
$\tilde{x}\leftarrow x+\sigma\alpha q^{(m)}$; \\*
$\widetilde{F}\leftarrow\left\Vert \widetilde{x}\right\Vert _{w1}$; \\*
\# Test the solution \\*
IF($\widetilde{F}-F\leq\theta$) THEN\{$x\leftarrow\tilde{x}$; $F\leftarrow\widetilde{F}$;\}\} \\*
\# Compute the new parameters \\*
$\theta\leftarrow\lambda\theta$; $\alpha\leftarrow\lambda\alpha$;\} \\*
\# Return the approximation of $x$ \\*
RETURN $x$; \\*

\section{Numerical Results}
\label{Numerical Results}

We have implemented the above SO algorithm in C on a Linux PC, using
the GNU Scientific Library \cite{key-14}. In Figure 2 we give several
examples of sparse signal recovery using the SO method. The non-zero
coefficients of the sparse signal $x\in\mathbb{R}^{M}$ are uniformly random 
generated in the interval $[-1,1]$. Also, the elements
of the measurement matrix $\varphi_{m}^{(n)}$ are drawn from the
Gaussian distribution $\Gamma(0,1)$. In these particular examples, the length
of the sparse signal $x\in\mathbb{R}^{M}$ and the number of measurements
were set to $M=100$, and respectively $N=50$. The initial and final
thresholds were set to $\theta_{i}=0.5$, and respectively $\theta_{f}=10^{-5}$.
Also we used $\alpha_{i}=1$, and $\lambda=0.95$, such that the number
of iterations are $T=\ln(\theta_{f}/\theta_{i})/\ln(\lambda)=300$.
The sparsity of the signal was varied as following: $K=20,25,30$.
One can see that by increasing $K$, and keeping the number of measurements
constant, the recovery error 
\begin{equation}
E=100\times\left\Vert x_{recovered}-x_{original}\right\Vert /\left\Vert x_{original}\right\Vert (\%)
\end{equation}
deteriorates: $E=8.532\cdot10^{-4}\%$
for $K=20$; $E=3.145\%$ for $K=25$; $E=11.329\%$ for $K=30$. This result is expected, 
since the fixed number of measurements cannot hold enough information about 
the increasing number of non-zero elements. Also, this result suggests that there is a phase 
transition in the error function, depending on the ratio between the sparsity 
parameter $K$ and the number of measurements $N$. 
\begin{figure}[h!]
\centering
\includegraphics[]{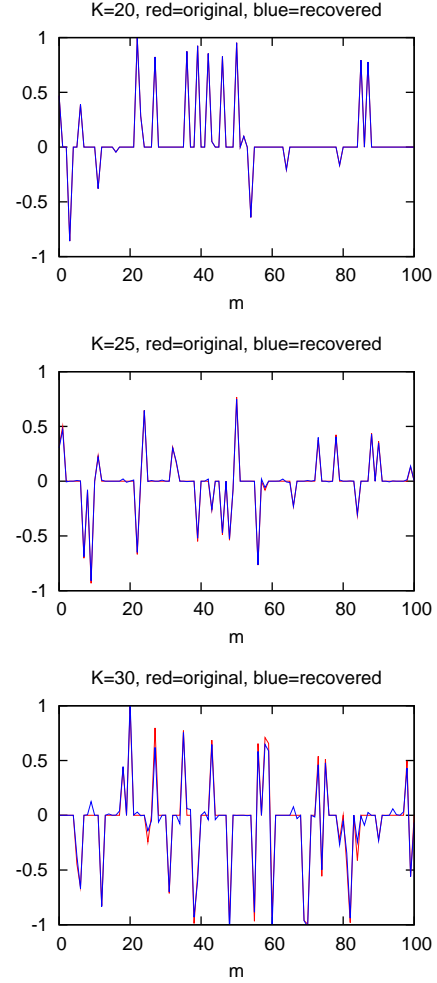}
\caption{Examples of sparse signal recovery.}
\label{Fig2}
\end{figure}

The phase transition is illustrated in Figure 3, 
where we give the relative recovery error $E=E(K,N)$, as a
function of the sparsity $K$ and the number of measurements $N$,
obtained by averaging over 100 instances for each value of $K$ and
$N$. One can see that there is a large area (the dark color) where the
method performs very well, and recovery is done with an error of less than 10\%. Also, the contour
lines, corresponding to a fixed error $E$, have a logarithmic dependence,
similar to the ones obtained with the BP approach.

It is interesting to note that the measured vector $y$ is used only
once in the whole recovery process, to find the admissible initial
solution, which satisfies the linear constraint system. After this
initial step, the algorithm simply perturbs the solution such that
the new candidate solution always satisfies the linear constraint
system. This approach reduces drastically the search space from $\mathbb{R}^{M}$
to $\mathbb{R}^{R}$, where $R=M-N$ is the rank of the null subspace
operator $\hat{Q}$.
As expected, by increasing the number of measurements $N$, the dimensionality
$R$ of the search space is reduced and the method will perform better.
\begin{figure}[h!]
\centering
\includegraphics[]{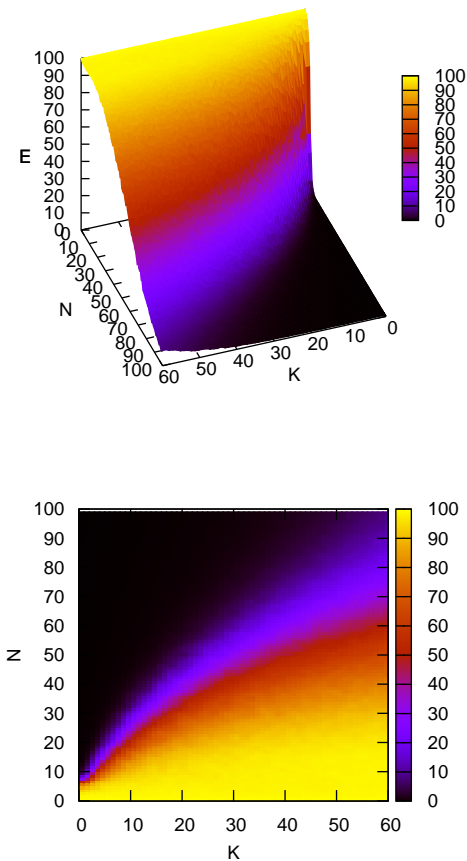}
\caption{The phase transition in the average recovery error, $E=E(K,N)$.}
\label{Fig3}
\end{figure}

\section{Conclusion}
\label{Conclusion}
We have presented a SO approach to the problem
of sparse signal recovery encountered in the CS framework. The considered
SO method has the advantage of a very easy implementation, comparing
to the highly complex BP standard approach, used in CS framework.
Thus, the proposed method is well suited for applications in which
a simple implementation, with an approximate recovery performance
of the original sparse signals, is acceptable. The objective function
of the SO method is defined as the product between the $\ell_{1}$
norm and the \textit{spectral entropy} of the candidate solution,
and it is equivalent with a weighted $\ell_{1}$ norm functional,
where the weights correspond to the \textit{self-information} of each
component of the signal. As a consequence of using such an objective
function, the convergence of the SO minimization method is improved,
since it focuses on entries where the weights are small, which by
definition correspond to the non-zero components of the sparse signal.

% >>>>>>>>>>>>>>>>>>>>>> Bibliography <<<<<<<<<<<<<<<<<<<<<<<<<<<<<<<<<<<<<
% A label is give for each bibitem.  Your paper references this label


\begin{thebibliography}{99}

% >>>>>>>>> Book examples <<<<<<<<<
%\bibitem{CarpenterBOOK} Carpenter, R.H.S., {\it Movements of the Eyes}, 2nd Edition, Pion Publishing, 1988.

%\bibitem{FranklinBOOK} Franklin, G.F., Powel, J.D., Workman, M.L., {\it Digital Control of Dynamic Systems}, Second Edition, Addison-Wesley, 1990.

% >>>>>>>>> Conference Proceedings Example <<<<<<<<<
%\bibitem{OhICRA1998} Oh, P.Y., Allen, P.K., ``Design a Partitioned Visual Feedback Controller,'' {\it IEEE Int Conf Robotics \& Automation}, Leuven, Belgium, pp. 1360-1365 5/98

% >>>>>>>>> Journal Example <<<<<<<<<<<<<<<<<<<<<<<<
%\bibitem{OhTRA2001} Oh, P.Y., Allen, P.K., ``Visual Servoing by Partitioning Degrees of Freedom,'' {\it IEEE Trans on Robotics \& Automation}, V17, N1, pp. 1-17, 2/01
 
 
\bibitem{key-1}Donoho, D., "Compressed Sensing", {\it IEEE Trans. Inf. Theory}, V52, pp. 1289-1306, 2006. 

\bibitem{key-2}Candes, E., Tao, T., "Near Optimal Signal Recovery
from Random Projections: Universal Encoding Strategies?", {\it IEEE Trans.
Inf. Theory}, V52, pp. 5406-5425, 2006. 

\bibitem{key-3}Candes, E., Wakin, M.B., "An Introduction To
Compressive Sampling", {\it IEEE Signal Proc. Mag.}, 25(2), pp. 21 - 30, 2008. 

\bibitem{key-4}Baraniuk, R., "Compressive Sensing", {\it IEEE Signal Proc. Mag.}, 24, pp. 118-121, 2007.

\bibitem{key-5}Romberg J., "Imaging via Compressive Sampling",
{\it IEEE Signal Proc. Mag.}, 25(2), pp. 14 - 20, 2008.

\bibitem{key-6}Compressive Sensing Resources at Rice University:
http://dsp.rice.edu/cs.

\bibitem{key-7}Andrecut, M., "Fast GPU Implementation of Sparse
Signal Recovery from Random Projections", {\it Eng. Lett.}, 17(3), pp. 151-158, 2009.

\bibitem{key-8}Andrecut, M., Kauffman, S. A., "On the Sparse
Reconstruction of Gene Networks", {\it J. Comput. Biol.}, V15, pp. 21-30, 2008.

\bibitem{key-9}Golub, G.H., van Loan, C. F., \textit{Matrix Computations},
3rd edition, Johns Hopkins Univ. Press, 1996.

\bibitem{key-10}Dueck, G., Scheuer, T., "Threshold Accepting:
A General Purpose Optimization Algorithm Appearing Superior to Simulated
Annealing", {\it J. Comput. Phys.}, 90, pp. 161-175, 1990.

\bibitem{key-11}Kirkpatrick, S., Gelatt, C. D. Jr, Vecchi, M. P.,
"Optimization by Simulated Annealing", {\it Science}, 220, pp. 671-680,
1983.

\bibitem{key-12}Cover, T.M., Thomas, J.A., \textit{Elements of Information
Theory}, Wiley, New York, 1991.

\bibitem{key-13}Powell, G.E., "A Spectral Entropy Method for
Distinguishing Regular and Irregular Motion of Hamiltonian Systems",
{\it J. Phys. A: Math. Gen.}, 12, pp. 2053-2071, 1979.

\bibitem{key-14}Galassi M. et al., \textit{GNU Scientific Library
Reference Manual - Third Edition}, Network Theory Limited, 2009.
\end{thebibliography}
\end{document}